# Unimodal part of the magnitude-frequency distribution of earthquakes according to the USGS catalog

A.V. Guglielmi

*Institute of Physics of the Earth RAS, Moscow, Russia,*
*guglielmi@mail.ru*

**Abstract**

In this paper, we used the Global Catalog of the National Earthquake Information Center US Geological Survey (NEIC USGS) for analysis of the magnitude-frequency distribution of earthquakes. We selected the unimodal part of the distribution and proposed an empirical formula that approximates this part in a wide range of magnitudes. The formula may be useful in studying the problems seismology, in which the expected effect is manifested in relatively weak earthquakes. The paper will be sent to the journal Geodynamics and Tectonophysics (ISSN 2078-502X).

**1. Introduction**

The global and regional catalogs of earthquakes contain a rich information about the Earth's seismicity [1–3]. In this paper, we use the Global Catalog of the National Earthquake Information Center US Geological Survey (NEIC USGS, https://earthquake.usgs.gov/earthquakes/). We will focus on the magnitude-frequency distribution of earthquakes.

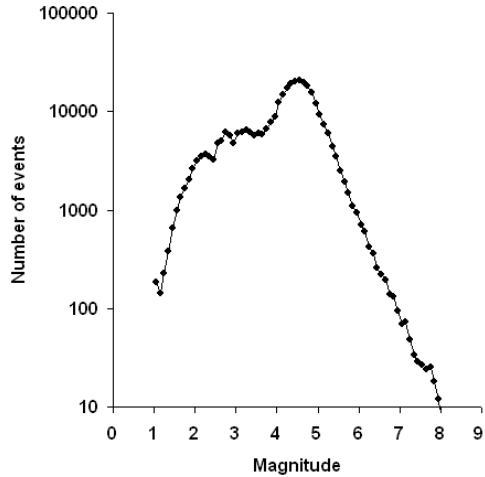

Fig. 1. Planetary earthquake distribution by magnitudes according to the USGS catalog.

Let's look at Figure 1. The figure shows the magnitude-frequency distribution of earthquakes recorded on Earth in the period from 1973 to 2014. We see that for M > 5 the distribution obeys the fundamental Gutenberg-Richter law [1, 4]

$$\lg N = a - b\mathrm{M}.\qquad(1)$$

The approximation of right branch of the magnitude-frequency distribution of earthquakes in the form (1) is widely used in seismology (e.g., see [1–9]). This part of the distribution is usually called representative.

Unrepresentative part of the distribution (approximately when M < 5) has a complex shape. The overall appearance of the distribution shown in Figure 1 tells us that we are most likely dealing with a bimodal distribution. In other words, our sample is probably a superposition of two different types of events.

We will try to highlight the unimodal part of the distribution. Then we select an empirical formula that approximates the unimodal part. Further, we will point out the literature, which uses not only the representative, but also unrepresentative part of the earthquake catalog. Finally, we briefly describe an interesting task for the solution of which the unrepresentative part of the USGS catalog was used.

## 2. Unimodal hyperbolic distribution

Let's try to separate the two types of distributions. Suppose that one type is formed in the Pacific "ring of fire" (PRF), and the second in the Alpine-Himalayan orogenic belt (AGB). In this preliminary study, we restrict ourselves to a fairly rough selection of events. Namely, we will select earthquakes in the following two areas: -90º – + 90º Lat, 80º – 180º Lon for PRF, and 0º – 50º Lat, 0º – 80º Lon for AGB.

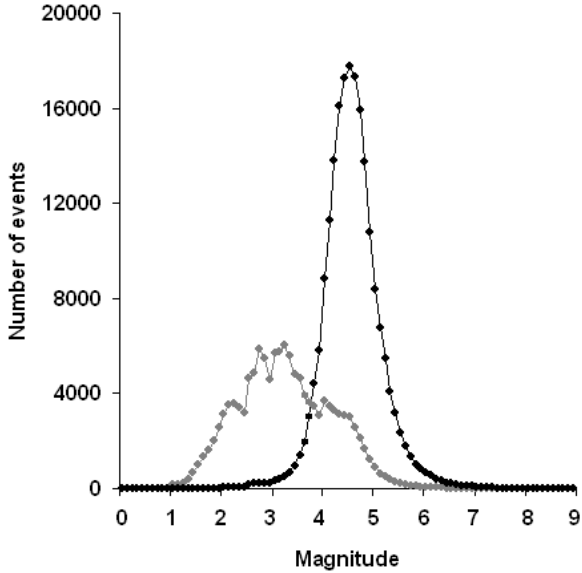

Fig. 2. Representation of Figure 1 in the form of two unimodal distributions.

The result is shown in Figure 2 by the black line for the PRF (201782 events) and the gray line for the AGB (129981 events). We see a clearly unimodal distribution of earthquakes in the PRF.

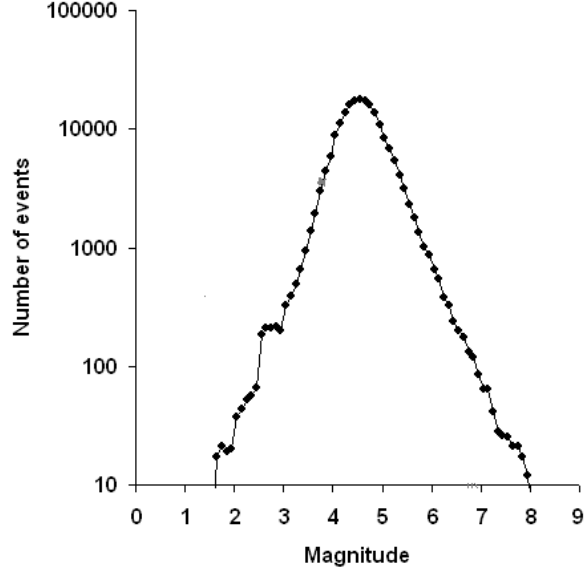

Fig. 3. Unimodal distribution on the semi-logarithmic scale.

Figure 3 shows the unimodal distribution in PRF on the semi-log scale. The picture resembles a wigwam, or an isosceles triangle. The top of the distribution is smoothed, so it's more correct to talk not about the triangle, but about the lower branch of an equilateral hyperbola. The right asymptote of the hyperbola is described by equation (1). It approximates the representative part of the distribution. The distribution over the entire range of magnitudes is described by the formula

$$\lg N = \alpha - \beta\sqrt{(M-M_0)^2 + \gamma^2} \; . \qquad (2)$$

For $M > M_0$ and $\gamma \to 0$ formula (2) coincides with the classical formula (1) up to the notation. Roughly $M_0 \approx 4.5$ and $\gamma \approx 0.1$. Thus, formula (1) is applicable for $M \geq 4.8$ with five percent accuracy.

B.I. Klain and author of this paper introduced the concept of the statistical sum Z of an earthquake ensemble [10]. For distribution (2) the statistical sum has the form

$$Z = \sum_j \exp[-\beta\sqrt{(M-M_0)^2 + \gamma^2}]. \qquad (3)$$

Here $M_j$ is the magnitude of the earthquake with number $j$. The entropy of ensemble is expressed through Z:

$$S = \ln Z - \beta \frac{d \ln Z}{d \beta}. \quad (4)$$

## 3. Discussion

Formulas (2)–(4) may be useful in studying those problems of seismology, in which it is desirable, and sometimes just necessary to take into account not only the representative, but also unrepresentative part of the earthquake catalog. Examples of this kind are described in the literature [7–12]. Sometimes, the expected effect is seen most clearly in the earthquakes of a relatively small magnitude. We present here one such example.

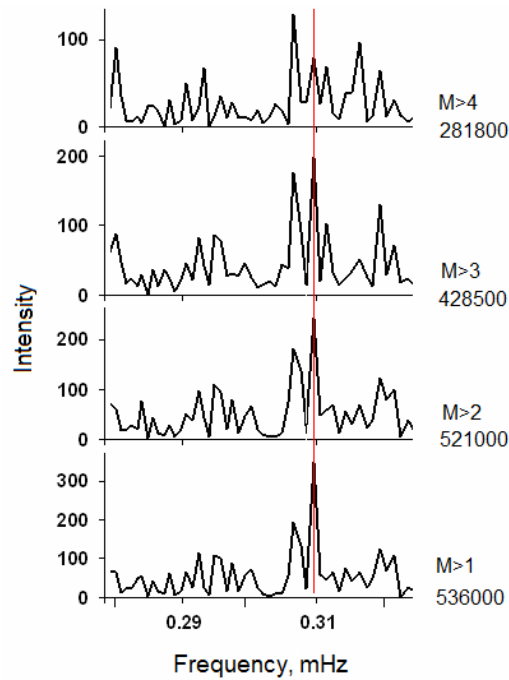

Fig. 4. The spectra of global seismicity for 1973–2010 [11]. The thin vertical line marks the frequency of the fundamental spheroidal mode of the free oscillations of the Earth.

Modulation of global seismicity by spheroidal oscillations of the Earth was discovered in the analysis of relatively weak earthquakes [11] (see also the review [9]). Figure 4 shows the seismicity spectra calculated from USGS data. The values of minimal magnitude M and the corresponding sizes of the samples are shown to the right of each spectrum. The thin vertical line marks the frequency of the fundamental spheroidal mode of oscillations. The peaks at the frequency of 0.31 MHz practically coincide with the frequency of the fundamental mode of oscillation. Apparently, Figure 4 is rather convincing evidence of the modulation of global seismicity by free oscillations of the Earth.

## 4. Conclusion

According to the USGS catalog, we selected the unimodal part of the magnitude-frequency distribution of earthquakes and found the empirical formula (2), which approximates this part of the distribution. At present, it is difficult to say how useful this result will be for seismology. In the future, we plan to use the formulas (2)–(4) for the study of variations in global seismicity.

*Acknowledgments*. I express my deep gratitude to B.I. Klain and O.D. Zotov for invaluable help in carrying out this work. I am sincerely grateful to A.D. Zavyalov for discussing earthquake physics. I thank the staff of USGS/NEIC for providing the catalogues of earthquakes. The work was supported by the Program No. 12 of the RAS Presidium, project of the Ministry of Education and Science of the Russian Federation KP19-270, project of the RFBR 18-05-00096, as well as the state assignment program of the IPhE RAS.